\documentclass{article}

\usepackage{PRIMEarxiv}

\usepackage[utf8]{inputenc} 
\usepackage[T1]{fontenc}    
\usepackage{hyperref}       
\usepackage{url}            
\usepackage{booktabs}       
\usepackage{amsmath}
\usepackage{amsfonts}       
\usepackage{nicefrac}       
\usepackage{microtype}      
\usepackage{lipsum}
\usepackage{fancyhdr}       
\usepackage{graphicx}       
\graphicspath{{media/}}     

\title{Developing a Ranking Problem Library (RPLIB) from a data-oriented perspective
\thanks{\textit{\underline{Citation}}: 
\textbf{Authors. Title. Pages.... DOI:000000/11111.}} 
}

\author{
  Paul E. Anderson, Brandon Tat, Charlie Ward \\
  Department of Computer Science and Software Engineering \\
  California Polytechnic State University \\
  San Luis Obispo, CA, USA\\
  \texttt{\{pander14, bjtat, cward11\}@calpoly.edu} \\
   \And
  Amy N. Langville, Kathryn E. Pedings-Behling \\
  Department of Mathematics \\
  College of Charleston \\
  Charleston, SC 29401, USA\\
  \texttt{\{langvillea, kepeding\}@cofc.edu} \\
}

\begin{document}
\maketitle

\begin{abstract}
We present an improved library for the ranking problem called RPLIB. RPLIB includes the following data and features. (1) Real and artificial datasets of both pairwise data (i.e., information about the ranking of pairs of items) and feature data (i.e., a vector of features about each item to be ranked). These datasets range in size (e.g., from small $n=10$ item datasets to large datasets with hundred of items), application (e.g., from sports to economic data), and source (e.g. real versus artificially generated to have particular structures). (2) RPLIB contains code for the most common ranking algorithms such as the linear ordering optimization method and the Massey method. (3) RPLIB also has the ability for users to contribute their own data, code, and algorithms. Each RPLIB dataset has an associated .JSON model card of additional information such as the number and set of optimal rankings, the optimal objective value, and corresponding figures.
\end{abstract}


\section{Introduction}
Constructing a ranking of items from a set of pairwise comparisons is a common problem with applications in a wide range of fields including sports, economics, social science, and many others. A \textit{ranking problem} can be defined as a dataset of items to be ranked. Such a dataset could include features of the items or pairwise comparisons between items. The current approach to producing rankings often focuses on selecting the single ranking best supported by the observed data but fails to communicate the ranking's uncertainty. In prior work \cite{anderson2021fairness,SIMODS,FODS,IGARDSTech3-moreunweighted,Cameron2020}, we introduced the idea of the rankability of a  ranking dataset. We tied rankability to the presence, number, and properties of multiple optimal solutions and thereby quantified a dataset's inherent ability to produce a meaningful ranking of its items.

Often progress in a particular field has been sparked by public datasets and libraries, e.g., ImageNet\cite{ImageNet}, the Linear Ordering Library\cite{LOLib}, and \href{Kaggle}{kaggle.com/}. For example, TSPLIB \cite{TSPLib} provides researchers of the traveling salesman problem (TSP) a unified place and problem instances by which to compare their algorithm's speed, scalability, and solutions and to compete for titles such as the largest TSP solved or fastest runtime. RPLIB, our Rankability Problem Library, was motivated by such libraries. 

We hope RPLIB will spark similar research for the ranking community and its instances will serve as comparative benchmarks. By describing RPLIB here, we also aim to encourage the ranking community to attack some of the open questions we pose in Section \ref{section:OpenQuestions}.

\section{Related work}
\begin{figure}
\centering
\includegraphics[height=6.2cm]{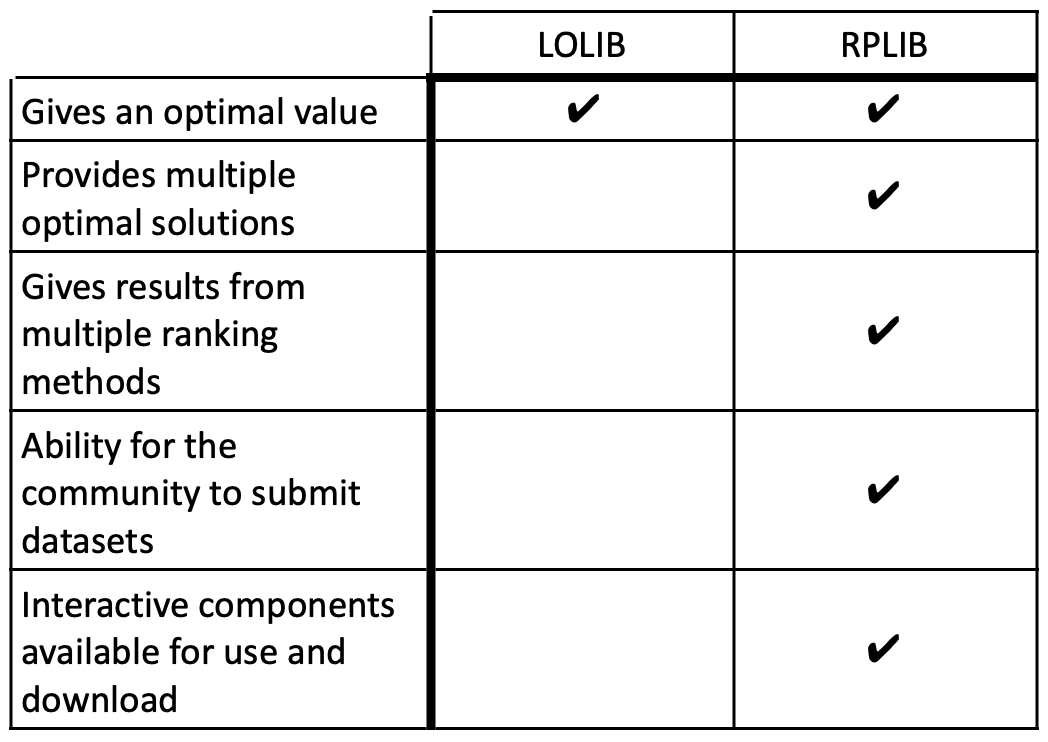}
\caption{A comparison of the features available in LOLIB versus those in RPLIB.}
\label{fig:LOvsRPLIB}
\end{figure}

RPLIB builds on the older existing ranking library LOLIB, the Linear Ordering Library \cite{LOLib}. LOLIB is an excellent repository of ranking data, but its datasets are static and in need of update. For example, an LOLIB text file contains just three pieces of information: the problem size, the optimal objective value for the linear ordering optimization, and one optimal ranking. In prior work \cite{FODS,anderson2021fairness}, we discovered that 98\% of the 50 instances in the LOLIB IO folder \cite{grotschel1984cutting} actually have more than one optimal ranking, yet just one is stored in the LOLIB text file. Unlike LOLIB, RPLIB algorithms produce and the library stores, in its novel JSON model card (Section \ref{section:modelcard}), all (or many if the user terminates the algorithm early) alternate optimal rankings associated with an instance. This is a novel contribution of RPLIB. Figure \ref{fig:LOvsRPLIB} summarizes several ways in which RPLIB extends LOLIB.

Another new contribution in this paper, in Section \ref{section:OpenQuestions}, is the extension of the definition of rankability from three measures to four measures. We use RPLIB instances and algorithms to demonstrate the additional benefits of this fourth rankability measure.

\section{RPLIB}

The RPLIB contains data, algorithms, code, and web applications for user contributions. These are described in the following subsections.

\subsection{Datasets}

RPLIB has three types of \textbf{unprocessed data}: (1) \emph{static real data} ranging from college basketball scores, features associated with U.S. colleges, organized and standardized versions of LOLIB instances, which include economic input-output instances as well as special artificial instances; (2) \emph{dynamic artificial data} that creates rankings with specific properties; and (3) \emph{user-contributed data}.

The RPLIB website contains additional details beyond those in this paper. See \url{https://igards.github.io/RPLib/}.

\subsubsection{Static Real Data}

RPLIB contains datasets from several real ranking applications.
\begin{itemize}
    \item Game data from seasons of U.S. college basketball, including the March Madness tournament. 
    \item economic input-output matrices from Japan's economic sectors for 1995 and 2005 \cite{Kondo2014}.
    \item Economic input-output matrices that were originally hosted within the linear ordering libraries, LOLIB and XLOLIB. These datasets have been updated and organized in RPLIB.
    \item Special collections from LOLIB and XLOLIB. These datasets were also cleaned and organized when moving them to RPLIB.
    \item US News \& World Report data on features associated with liberal arts colleges.
\end{itemize}

\begin{figure}
\centering
\includegraphics[height=2.5cm]{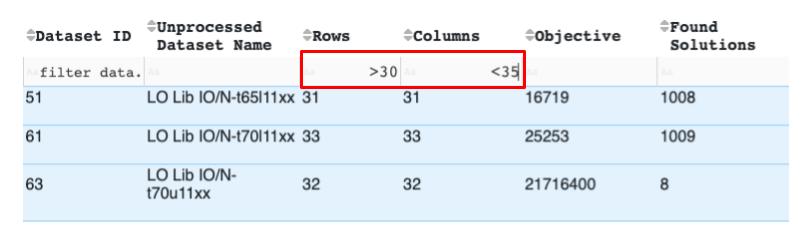}
\caption{This screenshot shows a powerful feature of RPLIB, the ability to filter the hundreds of problem instances to find instances satisfying certain conditions. This example  finds all instances from the LOLIB collection that have $30 < n< 35$.}
\label{fig:FilteringExample1}
\end{figure}

\begin{figure}
\centering
\includegraphics[height=5cm]{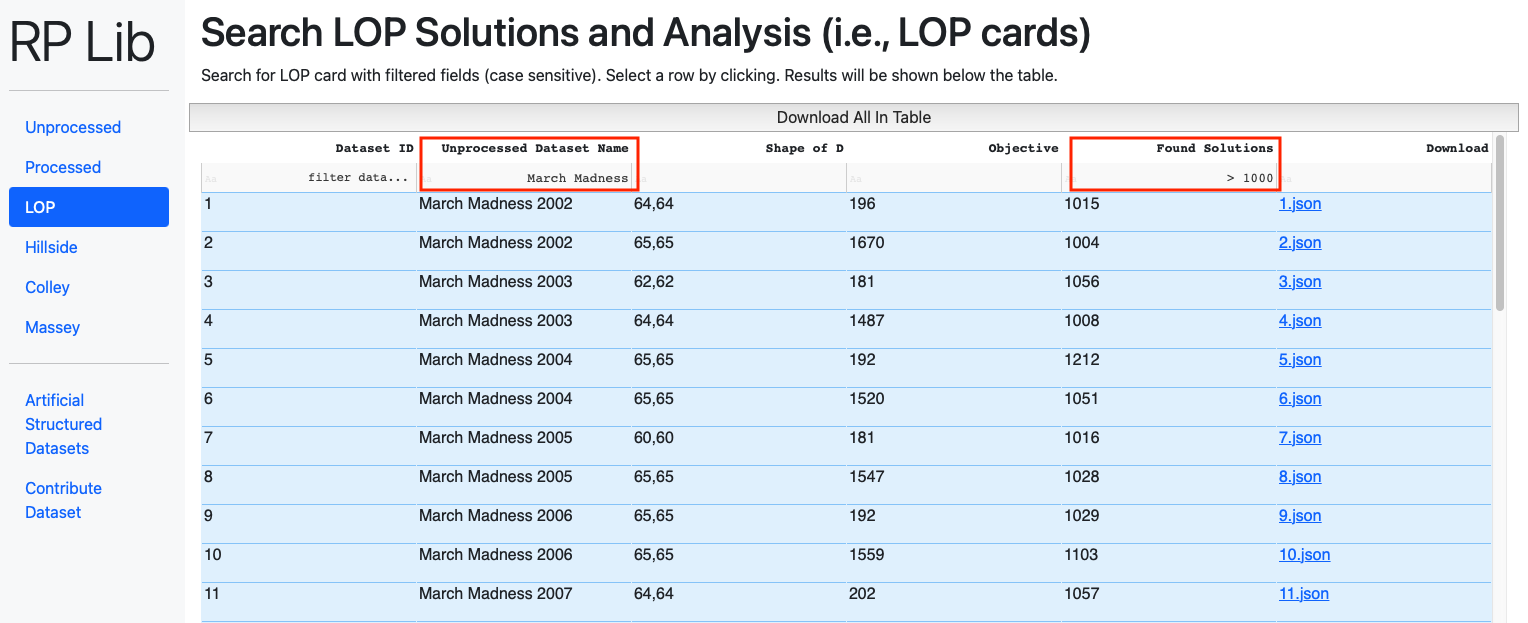}
\caption{This screenshot shows all March Madness instances that contain greater than 1000 optimal rankings.}
\label{fig:FilteringExample2}
\end{figure}

This collection can be filtered, searched, and exported for research purposes. For example, to find within RPLIB's hundreds of instances, instances from LOLIB where the problem size $n$ is between $30 \leq n \leq 35$, use the filter command as shown in Figure \ref{fig:FilteringExample1}.
To find all instances pertaining to March Madness basketball that contain greater than 1000 optimal rankings, use the filter command as shown in Figure \ref{fig:FilteringExample2}.

\subsubsection{Dynamic Artificial Data}

RPLIB also has the ability for users to \emph{generate} artificial instances that have desired structure. An RPLIB user has the ability to tune various parameters, such as the size of the instance or the amount of noise added to a matrix. These artificial datasets are useful for theoretical studies, particularly for the open questions posed in Section \ref{section:OpenQuestions}. Clicking on the ``Artificial Structured Datasets" tab on the left menu of the RPLIB homepage opens a Google Colaboratory Notebook.

\begin{figure}
\centering
\includegraphics[height=7.2cm]{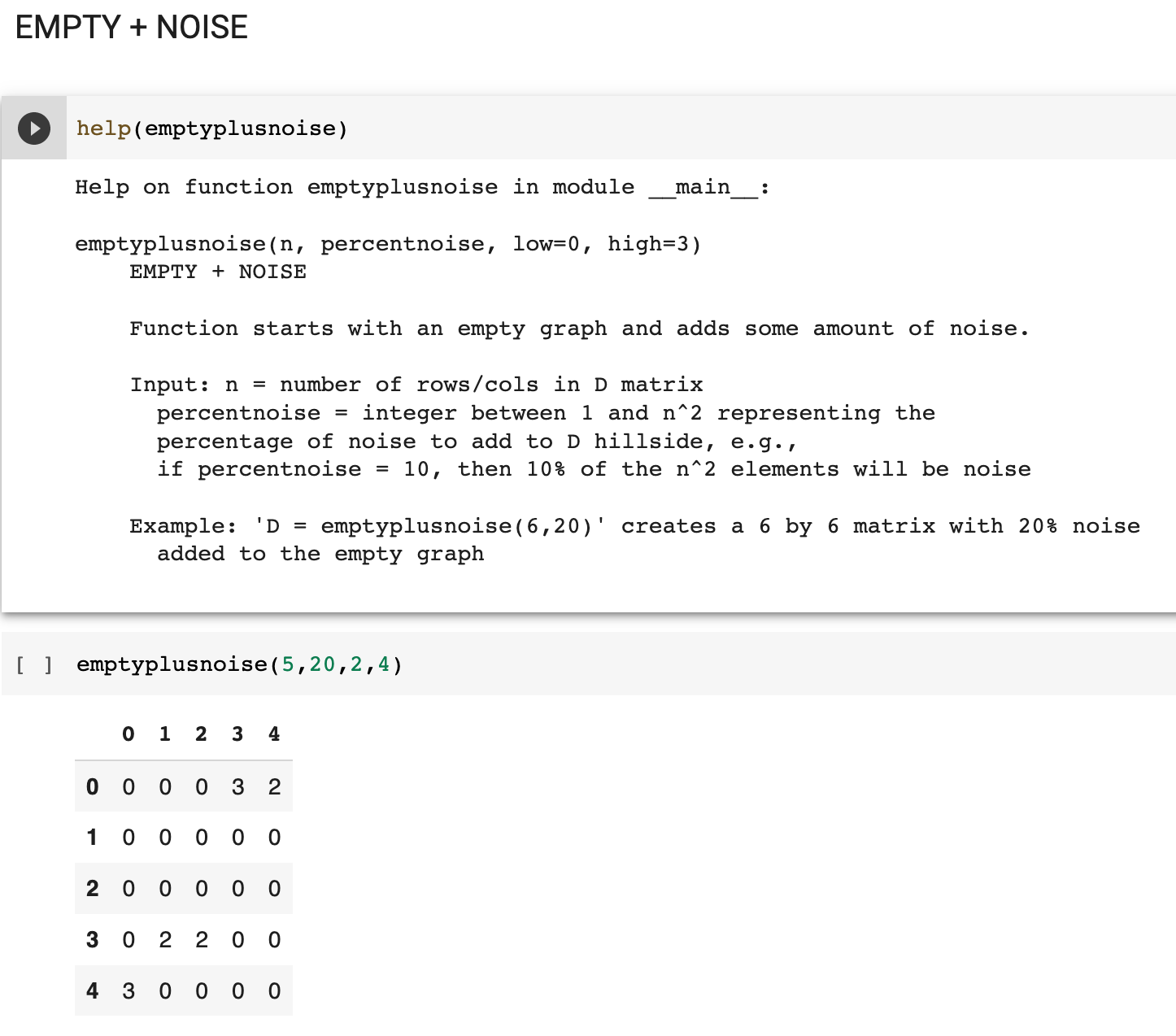}
\caption{The screenshot above shows how to create an artificial dataset in the Colab Notebook section of RPLIB. Specifically, this creates a $5 \times 5$ empty matrix where 20\% of the data is replaced with random entries between the lowerbound of 2 and the upperbound of 4.}
\label{fig:Colabexample}
\end{figure}

Figure \ref{fig:Colabexample} below is screenshot from the RPLIB Google Colaboratory Notebook. In this example, the command \texttt{emptyplusnoise(5,20,2,4)} creates a $5 \times 5$ empty matrix where 20\% of the data is replaced with random entries between the lowerbound of 2 and the upperbound of 4.

Currently the following structures of artificial data are available. These artificial datasets were first described in our prior work on rankability \cite{SIMODS}.
\begin{itemize}
    \item empty matrix (i.e., matrix of all zeros) 
    \item empty matrix plus a matrix with  random noise 
    \item a completely connected matrix
    \item a completely connected matrix minus a matrix with random noise
    \item perfect ranking matrix in hillside form plus random noise
    \item perfect ranking matrix in dominance form plus random noise
    \item cyclic matrix
    \item matrix structured so that it has $c!$ multiple optimal rankings with variations in the top, middle, or bottom of the ranking. See Section \ref{section:beta}.
    \item ability to turn weighted matrices into unweighted matrices
    \item ability to add or remove a percentage of entries in the matrix (i.e., edges in the associated graph)
    \item ability to generate a $\b D$ matrix by simulating individual games with a probability of upset
\end{itemize}

\subsubsection{User-contributed Data}

\begin{figure}
\centering
\includegraphics[height=8.2cm]{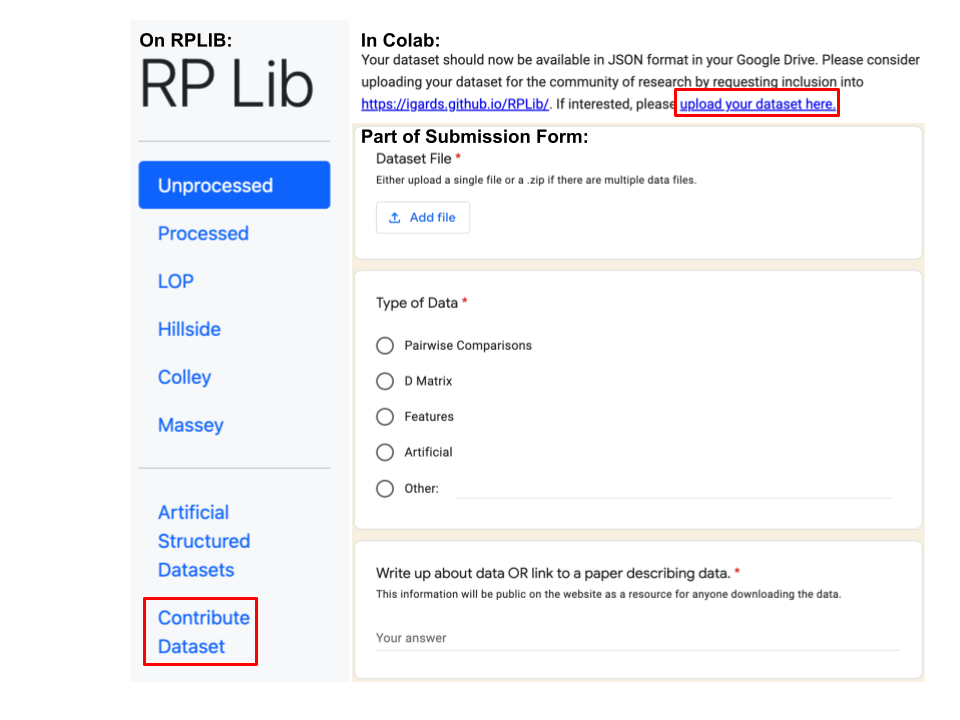}
\caption{There are two avenues for user to submit datasets. The left of the figure shows the submission link directly on the RPLIB website. The top of the figure shows a submission link within the Colab Notebook where users can create and submit their own artificial datasets. The remainder of the figure is a screenshot of a portion of the submission form.}
\label{fig:UserContributedData}
\end{figure}

RPLIB makes it easy for users to contribute their own data, either real static data or artificial structured data. There are two ways to make a contribution. (1) An RPLIB user can follow the ``Contribute Dataset" link on the RPLIB homepage, which opens a form that prompts the user for their name and email, the name and description of the dataset, the type of data, and the data itself. Contributions will be reviewed by the RPLIB team prior to being added to the library. (2) Alternatively, RPLIB users can create artificial datasets via RPLIB's Colab tool and then follow the link within that tool to directly submit their generated data. See Figure \ref{fig:UserContributedData}.

Ranking research has continued for decades, whereas rankability research has been recent. We invite researchers from these communities to add methods, data, and code to RPLIB. For example, rankability researchers from the U.S. and Europe have created interesting artificial datasets and new methods \cite{Cameron2020,McJamesRankability}.

\subsection{Dataset Processing}

\begin{figure}
\centering
\includegraphics[height=10cm]{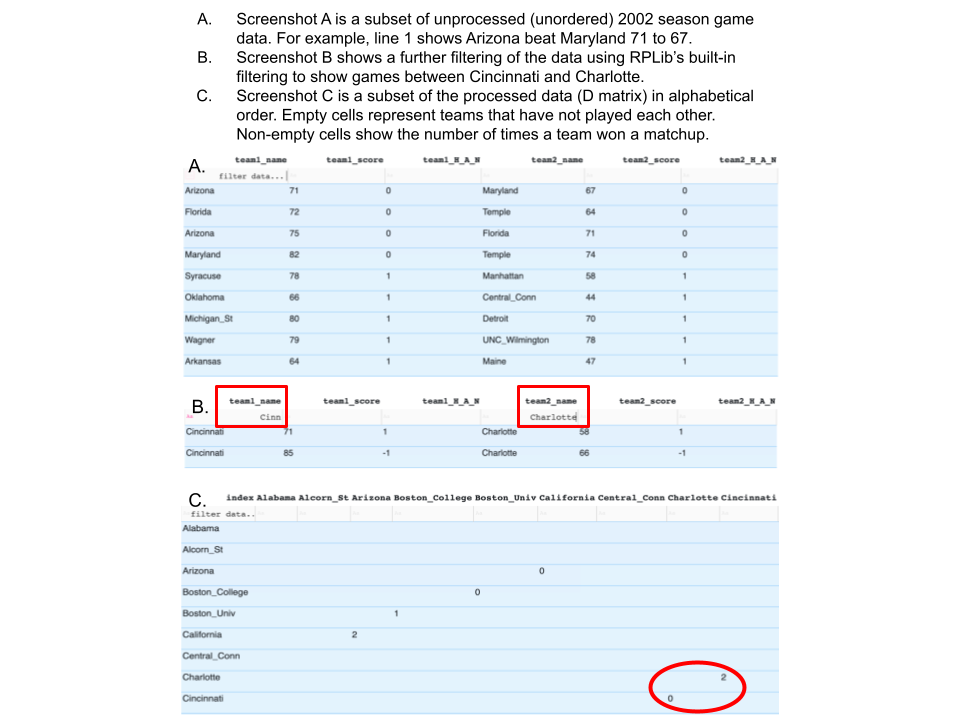}
\caption{These screenshots are taken directly from RPLIB. Screenshot A is the \emph{unprocessed} data for the 2002 March Madness basketball season. Screenshot B shows a powerful feature of RPLIB--the ability to search and filter data. In this example, we filtered the unprocessed data to find all instances of games between Cincinnati and Charlotte. Screenshot C is a subset of the (alphabetically ordered) \emph{processed} data, which creates a square matrix of pairwise comparisons. Empty cells mean the teams have not played each other. Non-empty cell $(i,j)$ is the number of times team $i$ beat team $j$ in that matchup.
}
\label{fig:processed+unprocesseddata}
\end{figure}

In this section, we use a few examples to explain the \textbf{Processed Data} section of RPLIB. Figure \ref{fig:processed+unprocesseddata} contains screenshots taken directly from RPLIB and highlights the difference between unprocessed and processed data. This example is for the 2002 season of March Madness basketball. Screenshot A is the \emph{unprocessed} data. Each row contains data for one game that season, showing the point score information. For example, row 1 indicates that Arizona scored 71 points, beating Maryland who scored 67 points. Screenshot B shows a powerful feature of RPLIB--the ability to search and filter data. In this example, we filtered this unprocessed instance to find all games between Cincinnati and Charlotte during that 2002 season. Contrast this with Screenshot C, which shows a subset of the (alphabetically ordered) \emph{processed} data, a square matrix of pairwise comparisons. An empty cell means the two teams have not played each other. Non-empty cell $(i, j)$ contains the number of times team $i$ beat team $j$ in that matchup.

\begin{figure}
\centering
\includegraphics[height=9cm]{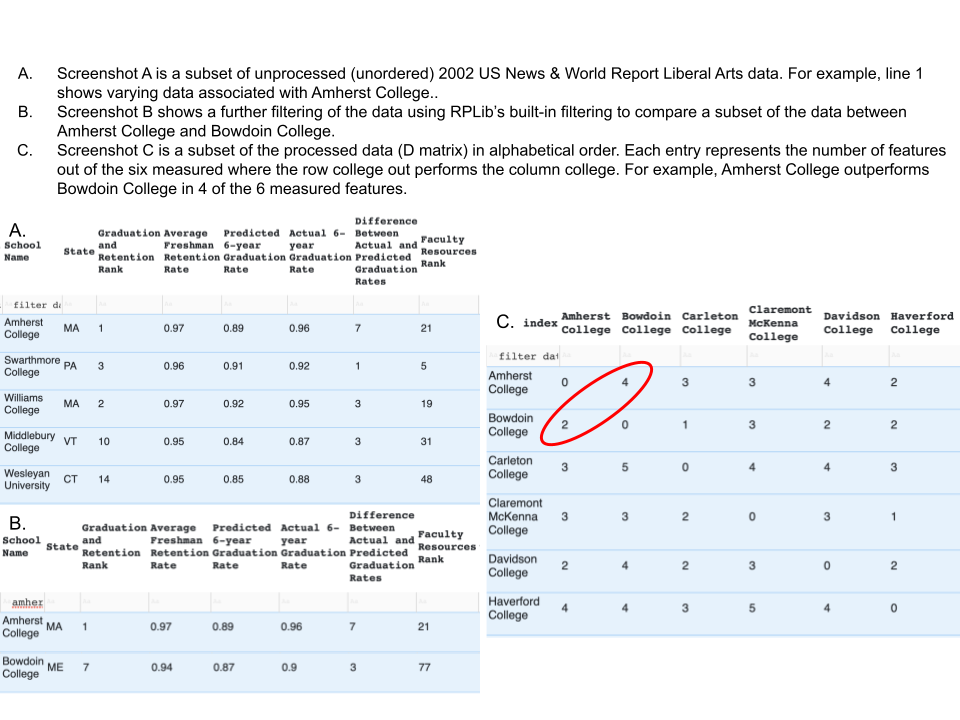}
\caption{These screenshots are taken directly from RPLIB. Screenshot A is the \emph{unprocessed} data for 2002 US News and World Report Liberal Arts College data. Screenshot B shows a powerful feature of RPLIB--the ability to search and filter data. In this example, we filtered the unprocessed data and compared the data from Amherst and Bowdoin Colleges. Screenshot C is a subset of the (alphabetically ordered) \emph{processed} data, which creates a square matrix of pairwise comparisons. Entry $(i,j)$ is the number of features out of the six total where college $i$ outperformed college $j$.
}
\label{fig:featuredataprocessing}
\end{figure}

Our next example is the US News \& World Report data, which is used to rank U.S. colleges.  RPLIB contains a subset of this data, specifically Liberal Arts colleges. Figure \ref{fig:featuredataprocessing} shows both the unprocessed and processed data.  Figure \ref{fig:featuredataprocessing}A is the \emph{unprocessed} data, which are data associated with various features (e.g., graduation rate, retention rate) for each college. Once again, the data is \emph{processed} to create a square matrix of pairwise comparisons, as shown in Figure \ref{fig:featuredataprocessing}B.

We invite the community to add new transformations that convert unprocessed data to processed data. This can be done with a standard
GitHub pull request. Once verified, such transformations can be run on all datasets through RPLIB's infrastructure for automated analysis.

\subsection{RPLIB Methods}
The library contains implementations of two classes of ranking methods:
\begin{enumerate}
\item Optimization: The linear ordering method \cite{Reinelt2011book} and its hillside variant \cite{Langville2011:MVR,SIMODS,FODS,IGARDSTech3-moreunweighted,HillsideAmountTechReport} are available. These methods output an optimal value, a collection of multiple optimal rankings (if any exist), the diameter of the set of multiple optimal rankings found, the solutions nearest to and farthest from the centroid of the optimal face, a matrix $\b X^*$ \cite{FODS} containing rank-ordered information about pairs of items, and a measure indicating the location of the indecision among the set of optimal rankings.
\item Linear Algebra: The Massey \cite{Massey1997} and Colley \cite{Colley2002} linear systems are available. These methods output the rating vector, its associated ranking vector, and a matrix $\b Y^*$ similar to the $\b X^*$ mentioned above, where $\b Y^*(i,j)$ gives information about the certainty that item $i$ is ranked above item $j$. The rating vector can be used to create a \textit{pseudo-optimal} set of rankings. 
\end{enumerate}

Users are able to download RPLIB code in order to run the optimization and linear algebra methods on their own data with their own machines. Alternatively, users can choose to upload their data using the \textbf{Contribute Dataset} option on RPLIB. The RPLIB team will process the data for contribution to the website. Depending on user feedback, future versions of RPLIB may include additional ranking methods, such as the Elo \cite{EloBook,LangvilleMeyerbook2} and learning to rank \cite{LearningtoRankBook} methods.

\subsection{RPLIB Outputs}

\subsubsection{RPLIB Model Card}
\label{section:modelcard}

Because ranking is a common task in machine learning, we adopt the standard documentation from that field, namely storing and visualizing the output as a \emph{model card} \cite{mitchell2019model}. An RPLIB instance contains a .JSON file and a Google Colab Notebook that generates images associated with that instance.  We intentionally chose JSON as the format for RPLIB data. JSON (JavaScript Object Notation) format is favored for its readability, nesting features, interoperability, and library support in modern programming languages. Because tools that convert JSON to other languages are common, our choice of JSON increases the usability of RPLIB. The JSON file contains some (or all) of the following items, which are displayed in our RPLIB tool as a model card. The Colab Notebook with images is described in the next section.
\begin{itemize}
    \item Dominance data. This can be either a square matrix $\b D$ of dominance relations between all pairs of items or a rectangular matrix of features for each item.
    \item Problem size $n$, i.e., the number of items to be ranked.
    \item Optimal objective value. As described in \cite{SIMODS}, many ranking and rankability methods optimize an objective function. 
    \item Optimal ranking(s).
    \item Diameter of the set of optimal rankings, i.e., the distance (e.g., Kendall tau) between the two farthest optimal rankings.
    \item Farthest (by Kendall tau) pair of optimal rankings.
    \item Closest (by Kendall tau) pair of distinct optimal rankings.
    \item Ranking closest to the centroid of all optimal rankings.
    \item Ranking farthest from the centroid of all optimal rankings.
    \item full set $P$ of all optimal rankings for some instances with small $n$.
    \item partial set $P$ of all known optimal rankings for larger instances.
\end{itemize}

\begin{figure}[h!]
\centering
\includegraphics[height=9cm]{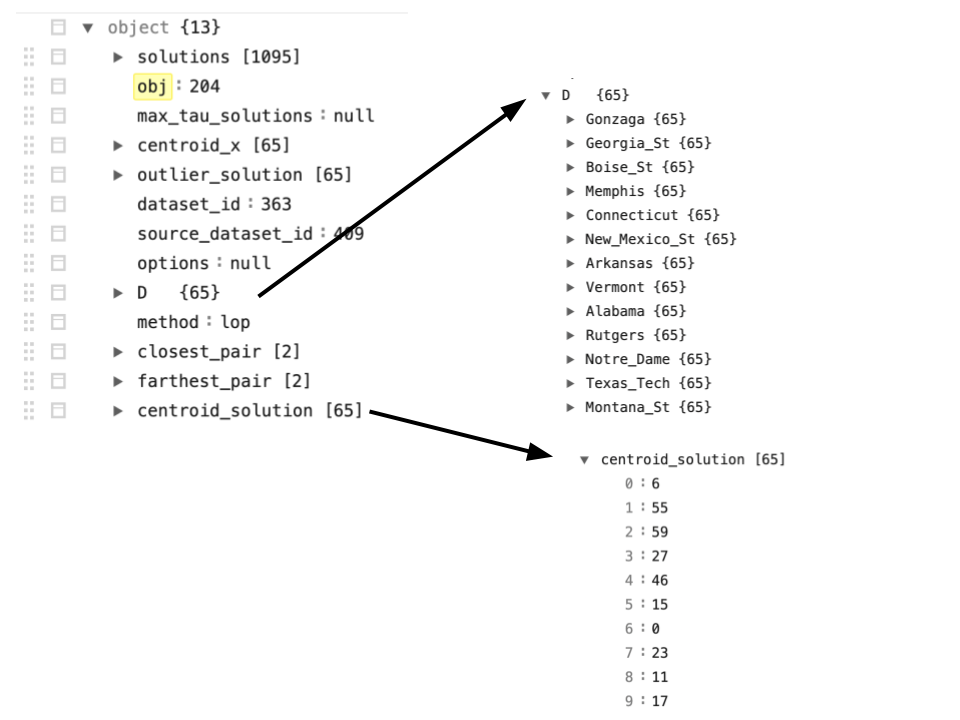}
\caption{A JSON Model Card. In this instance with dataset\_id 363, there are 1,095 optimal rankings with an objective value of 204. Clicking on a JSON item that has an arrow, such as D, expands to show more information. D is a 65 x 65 matrix. The first row corresponds to the team labeled Gonzaga and this can be further expanded to reveal the data for the 65 entries on this row of the D matrix. Similarly, expanding the centroid\_solution displays a vector of 65 elements, showing that the team with ID 0 is in rank position 6, team with ID 1 is in rank position 55, and so on.}
\label{fig:JSONModelCard}
\end{figure}

\subsubsection{RPLIB Figure Outputs}

Each RPLIB instance can be loaded in a Google Colab Notebook (see the top left of Figure \ref{fig:JSONModelCard}) and associated methods run, which generate associated measures (top right of Fig. \ref{fig:JSONModelCard}), and figures, such as a pixel plot of nonzero elements in a reordered $\b X(\b r, \b r)^*$ matrix and spaghetti plots that visually compare two rankings (see the bottom of Figure \ref{fig:JSONModelCard}).

\begin{figure}[h!]
\centering
\includegraphics[height=10.7cm]{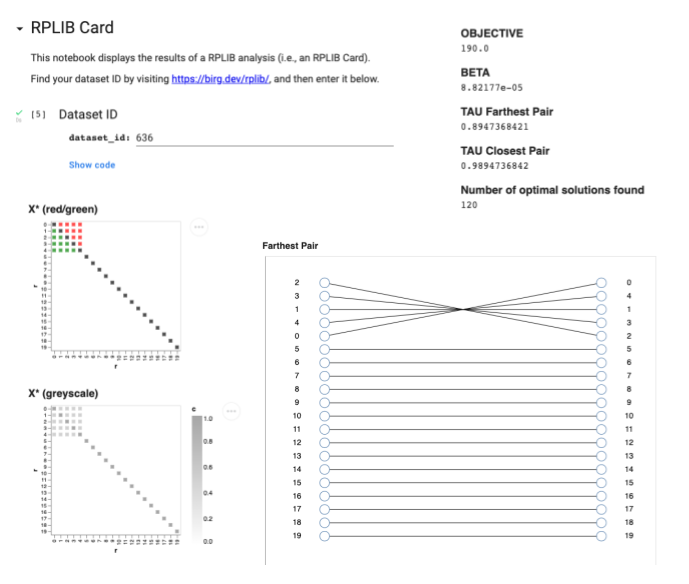}
\caption{A Google Colab Notebook provides the ability for users to generate and save any measures and images associated with a data instance. In this example, for the instance with ID 636, we show the pixel plots $\b X(\b r, \b r)^*$ matrix and a spaghetti plot between the farthest pair of optimal rankings.
}
\label{fig:JSONModelCard}
\end{figure}

\section{Example Use of RPLIB: refining the definition of rankability}
\label{section:beta}

\begin{figure}[h!]
\centering
\includegraphics[height=7cm]{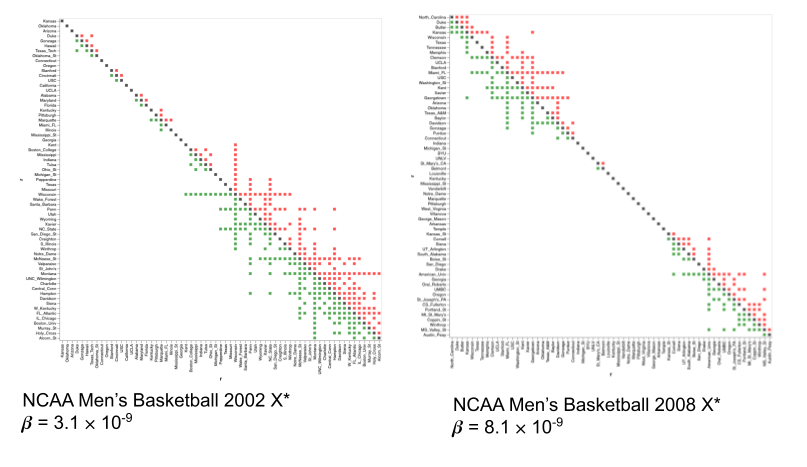}
\caption{$\b X^*$ matrices for two years of March Madness, 2002 (left) and 2008 (right).}
\label{fig:MM}
\end{figure}

In prior work, Anderson et al. define rankability $r$ as a function of three measures, $k$, the distance of $\b D$ from a perfectly rankable data matrix, $|P|$, the number of optimal rankings, and $\tau$, the diameter (i.e., the distance between the two farthest rankings) of the set of optimal rankings \cite{anderson2021fairness,SIMODS,FODS,IGARDSTech3-moreunweighted,HillsideAmountTechReport}.  Mathematically, they define $r=f(k, |P|, \tau)$ and show that rankability $r$ is correlated with predictability on the ranking task of predicting outcomes in the March Madness college basketball tournament. In other words, March Madness outcomes are straightforward and follow the rankings in more rankable years and are wild for less rankable years. Figure \ref{fig:MM} shows the $\b X^*$ matrices for two years of March Madness, 2002 and 2008. This figure plots the fractional elements in $\b X^*$, which measure the locations of disagreements among the optimal rankings in the set $P$ of optimal rankings. Notice that for 2002, there is little disagreement about the teams that belong at the top of the ranking, while there is much more disagreement about the bottom of the ranking.  The 2008 year shows more disagreement in the top of the ranking, certainty in the middle of the ranking, and some disagreement in the bottom. With respect to March Madness, we expect the 2008 year to have some upsets in later rounds of the tournament, when two top teams face each other.  In this section, \emph{we quantify the uncertainty in locations of the ranking by creating a fourth rankability measure, called $\beta$, which we define and describe in the remainder of this section.}

RPLIB provides a standardized approach for generating and archiving structured artificial data. We demonstrate the importance of archiving structured artificial data for the research community and the utility of RPLIB by expanding the definition of rankability from three measures to four. Mathematically, we test whether the following expanded definition of rankability is useful: $r=f(k, |P|, \tau, \beta)$, where $\beta$ is a measure of the location of indecision in multiple optimal rankings. In most ranking applications, one is most interested in the top-$k$ part of the ranking. 

\begin{figure*}
    \centering
    \includegraphics[width=10cm]{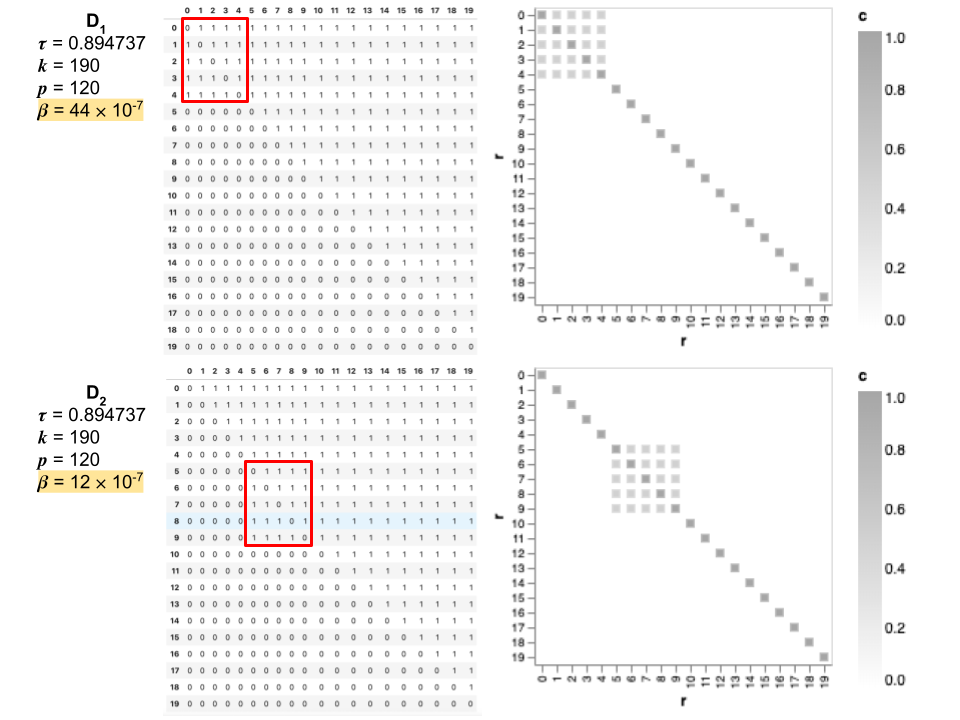}
    \includegraphics[width=10cm]{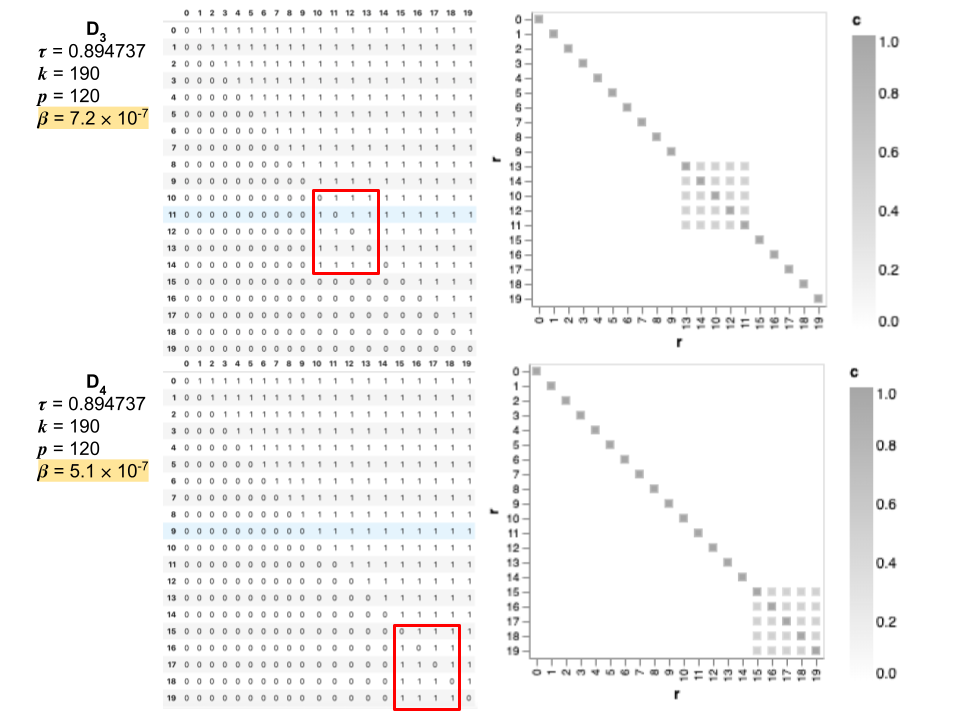}
    \caption{These four $\b D$ matrices (left) and their corresponding $\b X^*$ matrices (right) have the same $k$, $p=|P|$, and $\tau$ scores, but different $\beta$ scores.}
    \label{fig:DMatrices}
\end{figure*}

Thus, we use RPLIB to generate four artificial matrices from the \texttt{Special} class, $\b D_1$, $\b D_2$, $\b D_3$, and $\b D_4$, designed to have particular structure. Users have the ability in the RPLIB Colab Notebook to create matrices in the \texttt{Special} class that have a desired number of multiple optimal solutions. For example, when the user selects beginning and ending row indices of 6 and 10, respectively, this function creates a dominance matrix that has $(10-6+1)!=5!=120$ multiple optimal rankings. In this way, we generated the four dominance matrices $\b D_1$, $\b D_2$, $\b D_3$, and $\b D_4$ and their corresponding $\b X^*$ matrices shown in Figure \ref{fig:DMatrices}. 

\begin{figure}[h!]
\centering
\includegraphics[height=6.2cm]{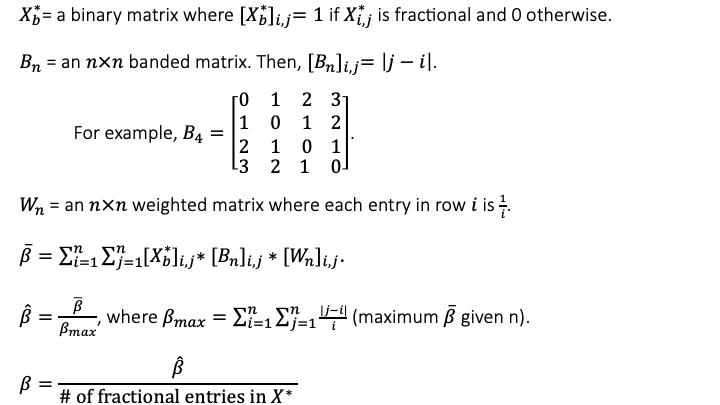}
\caption{Definition of $\beta$ for a matrix of size $n$}
\label{fig:DefofBeta}
\end{figure}

Ranking researchers prefer indecision, if it occurs, to appear toward the bottom of the ranking. This distinction is not captured by the existing rankability measures $k$, $p$, and $\tau$. Thus, we propose a fourth rankability measure $\beta$, defined in Figure \ref{fig:DefofBeta}, which uses the number and location of fractional entries of the $X^*$ matrix from the linear ordering model to penalize fractional entries that are further away from the diagonal and lower in the ranking.  Errors, ambiguity, or indecision in the bottom of rankings are less concerning than the same thing at the top of the ranking; the weighted definition of $\beta$ takes this into account.

This new rankability measure works as intended. Remember that lower scores are better. See the $\beta$ scores for $\b D_1$, $\b D_2$, $\b D_3$, and $\b D_4$ in Figure \ref{fig:DMatrices}. See Figure \ref{fig:MM} for the $\beta$ scores for the 2002 and 2008 years of March Madness. Thus, $\beta$ is capturing something that the other measures do not.

As shown in Figure \ref{fig:DMatrices}, the three prior metrics ($k$, $p$, $\tau$) do not capture the location of the multiple optimal solutions within the rankings (i.e., they do not capture position specific ambiguity in the rankings). Our proposed metric $\beta$ does take the position specific information into account. 

\section{Open Questions}
\label{section:OpenQuestions}

Our hope is that RPLIB will facilitate the study of open questions in the field of ranking. A few open questions follow. 

\begin{itemize}
\item \emph{Instance Difficulty.} What makes an instance of the ranking problem difficult, i.e., which instances require a long time to find one (or all) optimal ranking(s)? Are there properties of the input matrix that indicate the difficulty of a ranking instance?  

\item \emph{Inverse Problem.} Another open ranking problem is the inverse problem. Suppose you are given an optimal objective value and set of optimal rankings. Does there exist an input matrix associated with this output data? Under what conditions is the input matrix unique?

\item \emph{Rankability Score.} Anderson et al. \cite{anderson2021fairness,SIMODS,FODS} define the related ranking problem, the rankability problem. We hope RPLIB will advance open questions in that new field. For example, Anderson et al. defined rankability as a function of several measures such as the four measures described in Section \ref{section:beta}. How to combine these measures to create a scalar rankability score is an open question.  

\item \emph{Improving Rankability.} What data can be collected to improve the rankability of a given dataset? 

\end{itemize}

\section{Future Work}
In version 2.0 of RPLIB, we plan an interactive tool that enables users to analyze their own data using algorithms from the library. We also plan an interactive tool for one particular dataset, the US News and Report data on rankings of colleges from their features. We envision a tool that parents, students, and guidance counselors can use to personalize rankings of colleges for an individual student based on their preference and weights of features.

\section{Conclusion}
RPLIB is a library for researchers of the ranking (or linear ordering) problem. The library is a searchable database of hundreds of instances of ranking problems from both real and artificial data with algorithms, code, and an interactive notebook. We invite the ranking community to use the library, give feedback for future versions, and contribute work.

\section*{Acknowledgements}
We thank California Polytechnic State University students Marisa Aquilina, Ethan Goldfarb, and Jackson Waschura for their contributions to RPLIB.

\bibliographystyle{unsrt}  
\bibliography{references}

\end{document}